\documentstyle[epsfig,12pt]{article}

\date{\ }

\newtheorem{definition}{Definition}
\newtheorem{theorem}{Theorem}
\newtheorem{proposition}{Proposition}
\newtheorem{corollary}{Corollary}
\newtheorem{lemma}{Lemma}

\newcommand{\wfss}{well-founded semantics }
\newcommand{\kks}{Kripke-Kleene semantics }
\newcommand{\sems}{semantics }
\newcommand{\wfs}{well-founded semantics}
\newcommand{\kk}{Kripke-Kleene semantics}
\newcommand{\sem}{semantics}

\newcommand{\lfp}{least fixpoint }
\newcommand{\gfp}{greatest fixpoint }
\newcommand{\ifif}{if and only if }
\newcommand{\fa}{$\cal F$}
\newcommand{\tr}{$\cal T$}
\newcommand{\un}{$\cal U$}
\newcommand{\inc}{$\cal I$}
\newcommand{\fas}{$\cal F$ }
\newcommand{\trs}{$\cal T$ }
\newcommand{\uns}{$\cal U$ }
\newcommand{\incs}{$\cal I$ }  
\newcommand{\cqfd}{\mbox{}\nolinebreak\hfill\rule{2mm}{2mm}\medbreak}

\begin{document}
\title{Computing and Comparing Semantics of Programs in Multi-valued Logics\footnote{A preliminary version of this paper appeared in the form of an extended abstract in the conference {\it Mathematical Foundations of Computer Science (MFCS'99)}} }

\author{Yann Loyer \and Nicolas Spyratos \and Daniel Stamate\\\\
Laboratoire de Recherche en Informatique, UMR 8623,\\Universit\'e de Paris Sud,
         Bat. 490,
        91405 Orsay\\ \{loyer,spyratos,daniel\}@lri.fr }

\maketitle
\vspace{-1.5cm}         
\begin{abstract}

 The different semantics that can be assigned to a logic program correspond to different assumptions made concerning the atoms whose logical values cannot be inferred from the rules. Thus, the well founded semantics corresponds to the assumption that every such atom is false, while the Kripke-Kleene semantics corresponds to the assumption that every such atom is unknown. In this paper, we propose to unify and extend this assumption-based approach by introducing parameterized semantics for logic programs. The parameter holds the value that one assumes for all atoms whose logical values cannot be inferred from the rules. We work within  multi-valued logic with bilattice structure, and we consider the class of logic programs defined by Fitting. 

Following Fitting's approach, we define a simple operator that allows us to compute the parameterized \sem, and to compare and combine semantics obtained for different values of the parameter. The semantics proposed by Fitting corresponds to the value false. We also show that our approach  captures and extends the usual semantics of conventional logic programs thereby unifying their computation.

\end{abstract}

\section{Introduction}

The different semantics that can be assigned  to a logic program correspond to different assumptions made concerning the atoms whose logical values cannot be inferred from the rules. For example, the well founded semantics corresponds to the assumption that every such atom is false (Closed World Assumption), while  the Kripke-Kleene semantics corresponds to the assumption that every such atom is unknown. In general, the usual \sems of logic programs are given in the context of  three-valued logics, and are of two kinds: those based on the stable models \cite{gl:stable,przy1,przy3}  or on the \wfss \cite{gelder2}, and those based on the \kks  \cite{fitting:kk}.

We refer to semantics of the first kind as $pessimistic$, in the sense that it privileges negative information: if in doubt, then assume false; and we refer to semantics of the second kind as $skeptical$, in the sense that it privileges neither negative nor positive  information: if in doubt, then assume nothing. To illustrate these semantics,  consider the following program:
          
$$
\cal P
\left\{
\begin{array}{lll}
           \mbox{charge(X)} &   \leftarrow  &   \neg  \mbox{innocent(X)}  \wedge \mbox{suspect(X)}\\ 
            \mbox{free}(X) &  \leftarrow  & \mbox{innocent(X)} \wedge \mbox{suspect(X)}\\
            \mbox{innocent(X)} &  \leftarrow  & \mbox{free}(X)\\
       \mbox{suspect(John)} &   \leftarrow  &
\end{array}
\right.
$$

The only assertion made in the program is that John is suspect, but we know nothing as to whether he is innocent.

If we follow the pessimistic approach, then we have to assume that John is not innocent, and we can infer that John must not be  freed, and  must be charged. If, on the other hand, we follow the skeptical approach, then we have to assume nothing about the innocence of  John, and we can infer nothing as to whether he must be freed or charged.

However, in the context of three-valued logic, one can envisage a third \sem, that we shall call $optimistic$: if in doubt, then assume true. If we follow this  approach, then we have to  assume that John is  innocent, and we can infer that John must be freed, and must not be charged.

Now, the optimistic approach can be seen as a counterpart of the pessimistic approach. To find a counterpart for the skeptical approach, one has to adopt a multi-valued logic. In such a logic, one can envisage an $inconsistent$ semantics: if in doubt, then assume both false and true. Table~1 summarizes the four possible \sems of $\cal P$, where \fa, \tr, \uns and \incs stand for false, true, unknown and inconsistent, respectively.

          \begin{center}
          \begin{tabular}{|l|c|c|c|c|}
          \hline
           Approach&suspect(John)&innocent(John)&free(John)&charge(John)\\
          \hline
          Pessimistic & \tr & \fa & \fa & \tr\\
          \hline
          Optimistic & \tr & \tr & \tr & \fa\\
          \hline
          Skeptical  & \tr & \un & \un & \un\\
          \hline
          Inconsistent & \tr & \inc & \inc & \inc\\
          \hline
\end{tabular}
\end{center}
\begin{center}
Table 1 - {\it The four possible \sems of $\cal P$}
\end{center}

In this paper, we define the semantics of a program $\cal P$ using a parameter $\alpha$ whose value can be any of the above four logical values. Once fixed, the value of $\alpha$ represents the ``default value'' for those atoms of $\cal P$ whose values cannot be inferred from the rules. We define a simple operator that allows us to compute this parameterized \sem, and also to compare and combine \sems obtained for different values of $\alpha$. We show that our semantics extends the semantics proposed by Fitting \cite{fitting}, and captures the usual semantics of conventional logic programs thereby unifying their computation. As a side-result, we propose a new semantics for logic programs, that can be roughly described as a ``compromise'' between pessimistic and optimistic \sem.

Motivation for this work comes from the area of knowledge acquisition, where contradictions  may occur during the process of collecting knowledge  from different experts. Indeed, in   multi-agent systems, different agents may give different answers to the same query. It is then important to be able to process the answers so as to extract the maximum of information on which the various agents agree, or to detect the items on which the agents give conflicting answers.

Motivation also comes from the area of  deductive databases. Updates leading to a certain degree of inconsistency should be allowed because inconsistency can lead to  useful information, especially within the framework of distributed databases. In particular,  Fuhr and R\"{o}lleke showed in \cite{fr97} that hypermedia retrieval requires the handling of inconsistent information.

The use of multi-valued logics is justified by the fact that it provides a more natural modeling framework for the application areas just mentioned. Moreover, as  Arieli and Avron showed in \cite{arieliavron98}, the use of four values is preferable to the use of three even for tasks that can in principle be handled using only three values. 

The remaining of the paper is organized as follows. In section 2, we recall very briefly definitions and notations from three-valued  and multi-valued logics, namely, stable models and well-founded semantics,  Kripke-Kleene semantics, Belnap's logic, bilattices and Fitting's programs. We then proceed, in section 3, to define our parameterized semantics of a  Fitting program~$\cal P$. This is done by defining a parameterized operator whose fixpoints  we call the $\alpha$-fixed models of $\cal P$. Our treatment in this section is inspired by \cite{fitting}. If the value of the parameter $\alpha$ is false, then the $\alpha$-fixed models correspond to  the stable models proposed by Fitting. We also present an algorithm for computing the $\alpha$-fixed semantics of $\cal P$. In section 4, we restrict our attention to conventional logic programs. We show that their $\alpha$-fixed models capture the three-valued stable models, the well-founded semantics, and the Kripke-Kleene semantics. We also  provide a comparative study of the $\alpha$-fixed models for the four values of the parameter $\alpha$, and propose a ``compromise'' between pessimistic and optimistic semantics that in certain cases may lead to the definition of a new semantics.  Section 5 contains concluding remarks and suggestions for further research.

\section{Preliminaries}

\subsection{Three-valued logics}

\subsubsection{Stable models and well founded semantics}

Gelfond and Lifschitz introduced the notion of stable model \cite{gl:stable}, in the framework of classical logic under the closed world assumption. This notion was then extended to three-valued logics and partial interpretations: Van Gelder, Ross and  Schlipf introduced the well-founded semantics \cite{gelder2}, and Przymusinski defined the three-valued stable models \cite{przy1}. In fact, as shown in \cite{przy3}, Przymusinski's extension  captures both the bi-valued stable models and the well-founded semantics.

In Przymusinski's approach, a conjunctive logic program is a set of clauses of the form  $A \longleftarrow B_1 \wedge ... \wedge B_n \wedge \neg C_1 \wedge ...\wedge  \neg C_m
$, where $B_1,...,B_n,C_1,...,C_m$ are atoms. In this context, a valuation is a mapping that assigns to each ground atom a truth value from the set   \{$false, unknown, true$\}. A valuation  can be extended to ground litterals and  conjunctions of ground litterals in the usual way. To define the stable models and \wfss of a program~$\cal P$, one uses the extended Gelfond-Lifschitz transformation $GL_{\cal P}$ \cite{przy1} which assigns to each valuation $v$ another valuation $GL_{\cal P}(v)$ defined as follows : 
\begin{enumerate}
\item Transform  $\cal P$ into a positive program  ${\cal P}_{/v}$ by replacing all negative literals by their values from  $v$.
\item Compute the least fixpoint of an immediate consequence operator $\Phi$ defined as follows :
\begin{itemize}
\item if the ground atom $A$ is not in the head of any rule of Inst-${\cal P}_{/v}$, then ${\Phi}_{{\cal P}_{/v}} (v)(A) = false$; here, Inst-${\cal P}_{/v}$ denotes the set of all instantiations of rules of ${\cal P}_{/v}$;
\item  if the rule ``$A \longleftarrow$'' occurs in Inst-${\cal P}_{/v}$, then ${\Phi}_{{\cal P}_{/v}} (v)(A) = true$;
\item  else  ${\Phi}_{{\cal P}_{/v}} (v)(A) = \bigvee \{v(B)~|~A \leftarrow B \in$ Inst-${\cal P}_{/v}$, where $\vee$ is the extension of classical disjunction defined by:

\vspace{-.15cm}
\begin{center}
\begin{tabular}{ccccc}
 $false$ &$\vee$&$ unknown$&$ =$&$ unknown;$\\
$  true$&$ \vee$&$ unknown$&$ =$&$ true;$\\
$ unknown$&$ \vee$&$ unknown$&$ =$&$ unknown.$
\end{tabular}
\end{center}
\end{itemize}
\end{enumerate}

The valuation $v$ is defined to be a three-valued stable model of $\cal P$ if  $GL_{\cal P}(v) = v$. The least three-valued stable model coincides with the well-founded semantics of $\cal P$, as defined by Van Gelder et als \cite{gelder2}. 

It follows from the definition of $\Phi$ above that this approach gives greater importance to negative information, so it is a pessimistic approach.\\

\subsubsection{Kripke-Kleene semantics} 

Working with three-valued logic, Fitting introduced the Kripke-Kleene semantics \cite{fitting:kk}. The program $\cal P$ has the same definition as for  stable models, but the operator  $\Phi$ is now defined as follows :

\newpage

 given  a valuation $v$ and  a ground atom $A$ in Inst-$\cal P$,
\begin{itemize}
\item if there is a rule in  Inst-$\cal P$ with head  $A$, and the truth value of the body under $v$ is  $true$, then ${\Phi}_{\cal P} (v)(A) = true$;
\item  if there is a rule in  Inst-$\cal P$ with head $A$, and for every rule in Inst-$\cal P$ with head $A$ the truth value of the body under $v$ is false, then  ${\Phi}_{\cal P} (v)(A) = false$;
\item else  ${\Phi}_{\cal P} (v)(A) = unknown$.
\end{itemize}

It follows that this approach gives greater importance to the lack of information since $unknown$ is assigned to the atoms whose logical values cannot be inferred from the rules, so it is a skeptical approach.

\subsection{Multi-valued logics}

\subsubsection{Belnap's logic}

In \cite{belnap:four}, Belnap defines  a logic called $\cal FOUR$ intended to deal with incomplete and inconsistent information. Belnap's logic uses four logical values, that we shall denote by \fa, \tr, \uns and \incs, i.e. $\cal FOUR$ = \{\fa, \tr, \un, \inc\}. These values can be compared using two orderings, the knowledge ordering and the truth ordering.

In the knowledge ordering, denoted by ${\leq}_k$, the four values are ordered as follows: \uns ${\leq}_k$ \fa, \uns ${\leq}_k$ \tr, \fas ${\leq}_k$ \inc, \trs ${\leq}_k$ \inc. Intuitively, according to this ordering, each value of $\cal FOUR$ is seen as a possible knowledge that one can have about the truth of a given statement. More precisely, this knowledge is expressed as a set of classical truth values that hold for that statement. Thus, \fas is seen as \{$false$\},  \trs is seen as \{$true$\}, \uns is seen as $\emptyset$ and \incs is seen as \{$false$, $true$\}. Following this viewpoint, the knowledge ordering is just the set inclusion ordering.

In the truth ordering, denoted by ${\leq}_t$, the four logical values are ordered as follows: \fas ${\leq}_t$ \un, \fas ${\leq}_t$ \inc, \uns ${\leq}_t$ \tr, \incs ${\leq}_t$ \tr.  Intuitively, according to this ordering, each value of $\cal FOUR$ is seen as the degree of truth  of a given statement. \uns and \incs are both less false than \fa, and less true than \tr, but \uns and \incs are not comparable.

The two orderings are represented in the double Hasse diagram of Figure~1.

\begin{figure}
\centering
       \setlength{\unitlength}{0.35pt}
        \input{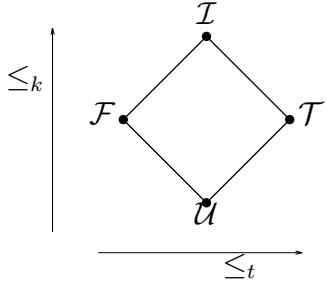}
        \caption{The logic {\sl FOUR}
        \label{four}}
\end{figure}

Each of the orderings $\leq_t$ and $\leq_k$ gives  $\cal FOUR$  a lattice structure. Meet and join under the truth ordering are denoted by $\wedge$ and $\vee$, and they are natural 
generalizations of the  usual notions of  conjunction and 
disjunction. In particular, \un $\wedge$\inc = \fas and \un $\vee$\inc = \tr. Under the knowledge ordering, meet and join are denoted by $\otimes$
and $\oplus$, and are called the $consensus$ and $gullibility$, respectively:
\begin{itemize}
\item $x
\otimes y$ represents the maximal information on which $x$ and  $y$  agree, whereas
\item $x \oplus
y$ adds the knowledge represented by  $x$ to that represented by  $y$.
\end{itemize}

 In particular, \fa $\otimes$\tr = \uns and \fa $\oplus$\tr = \inc.

There is a natural  notion of $negation$ in the truth ordering denoted by~$\neg$, for which we have: $\neg$ \tr = \fa, $\neg$ \fa = \tr, $\neg$ \un = \un, $\neg$ \inc = \inc. There is a similar  notion
for the knowledge ordering, called  $conflation$,
denoted by  -, for which: - \un = \inc, - \inc = \un, - \fa = \fa, - \tr = \tr.

The operations $\vee, \wedge, \neg$ restricted to the values \trs and \fas are those of classical logic, and if we add to these operations and values the value \un, then they are those of Kleene's strong three-valued logic.\\

\subsubsection{Bilattices}

In \cite{fitting:bilat,Mes97},  
bilattices are used as truth-value spaces for integration of
information coming from different sources.
The bilattice approach is a basic contribution to
many-valued logics. Bilattices and their derived sublogics
are useful in expressing uncertainty and inconsistency
in logic programming and databases \cite{arieliavron98,fitting,MPS97,SS97}. 
 The simplest non-trivial 
bilattice is called {\sl FOUR}, and it is basically Belnap's
four-valued logic \cite{belnap:four}.

\begin{definition}
A bilattice is a triple $\langle {\cal B},\leq_{t},\leq_{k}\rangle$, 
where $\cal{B}$ 
is a nonempty set and $\leq_t$, $\leq_k$ are each a partial ordering 
giving 
$\cal{B}$ the structure of a lattice with a top and a bottom.
\end{definition}

In a bilattice  $\langle {\cal B},\leq_{t},\leq_{k}\rangle$,
meet and join under $\leq_t$ are denoted $\vee$ and $\wedge$, and
meet and join under $\leq_k$ are denoted $\oplus$ and $\otimes$. Top and
bottom under $\leq_t$ are denoted ${\cal T}$ and ${\cal F}$, and top
and
bottom under $\leq_k$ are denoted $\cal I$ and $\cal U$. 
If the bilattice is complete with respect to both orderings, infinitary meet and join under $\leq_t$ are denoted $\bigvee$ and $\bigwedge$, and infinitary 
meet and join under $\leq_k$ are denoted $\bigoplus$ and~$\bigotimes$.

\begin{definition}
A bilattice  $\langle {\cal B},\leq_{t},\leq_{k}\rangle$ is called  distributive if all 12 distributive laws connecting $\vee$, $\wedge$, $\oplus$ and $\otimes$ hold. It is called infinitely distributive if it is a complete bilattice  in which all infinitary, as well as finitary, distributive laws  hold. 
\end{definition}

An example of a distributive law is $x\otimes(y\vee z)=(x\otimes y) \vee (x\otimes z)$. An example of an infinitary distributive law is $x\otimes \bigvee \{y_i | i \in S\} = \bigvee \{x \otimes y_i | i \in S\}$.

\begin{definition}
A  bilattice  $\langle {\cal B},\leq_{t},\leq_{k}\rangle$ satisfies the interlacing conditions if each of the operations $\vee$, $\wedge$, $\oplus$ and $\otimes$ is monotone with respect to both orderings. If the bilattice is complete, it satisfies the infinitary interlacing conditions if each of the infinitary meet and join is monotone with respect to both orderings.
\end{definition}

An example of an interlacing condition is:  $x_1 \leq_t y_1$ and $x_2 \leq_t y_2$ implies $x_1 \otimes x_2 \leq_t y_1 \otimes y_2$. An example of an infinitary interlacing condition is: $x_i \leq_t y_i$ for all $i \in S$ implies $\bigoplus\{x_i | i \in S\} \leq_t \bigoplus\{y_i | i \in S\}$. A distributive bilattice satisfies the interlacing conditions.

$\cal FOUR$ is an infinitary distributive bilattice which satisfies the infinitary interlacing laws. A bilattice is said to be {\em nontrivial} if the bilattice 
{\sl FOUR} can be isomorphically embedded in it.

A way for constructing a bilattice is proposed in \cite{ginsberg:mv val}. Consider two lattices $\langle L_1, \leq_1 \rangle$ and $\langle L_2, \leq_2 \rangle$. We can see $L_1$ as the set of values used for representing the degree of belief (evidence, confidence, etc.) of an information and $L_2$ as the set of values used for representing the degree of doubt (counter-evidence, lack of confidence, etc.) of the information. 

The structure $\langle L_1\times L_2,\leq_t,\leq_k\rangle$ where:
\begin{itemize}
\item $\langle x, y\rangle\leq_t\langle z,w\rangle$ iff
$x\leq z$ and $w\leq y$,\\
($\langle x, y\rangle\;\wedge\; \langle z,w\rangle$ =
$\langle min(x,z), max(y,w)\rangle$), and
\item $\langle x, y\rangle\leq_k\langle z,w\rangle$ iff
$x\leq z$ and $y\leq w$\\
($\langle x, y\rangle\;\otimes\; \langle z,w\rangle$ =
$\langle min(x,z), min(y,w)\rangle$)
\end{itemize} is a bilattice satisfying the interlacing conditions; it also satisfies the infinitary interlacing conditions if $L_1$ and $L_2$ are complete. Moreover, it is infinitely distributive if $L_1$ and $L_2$ are complete and infinitely distributive.

By abuse of notation we will sometimes talk about the bilattice
$\cal B$ when the orders are  irrelevant or understood from the context. 
From now on, we assume  that the bilattices we use are infinitely distributive, satisfy the infinitary interlacing conditions and have a negation
 unless explicitly stated otherwise.

\subsubsection{Fitting  programs}

Conventional logic programming has the set \{\fa, \tr\} as its intended space of truth values, but since not every query may produce an answer, partial models are often allowed (i.e. \uns is added). If we want to deal with inconsistency  as well, then \incs must be added. Thus Fitting asserts that $\cal FOUR$ can be thought as the ``home'' of ordinary logic programming and extends the notion of logic program so that a bilattice $\cal B$ other than $\cal FOUR$ can be thought of as the space of truth values.

\begin{definition}{\bf (Fitting program)}

\begin{itemize}
  \item  A formula is an expression built up from literals and elements of $\cal B$, using $ \wedge , \vee , \otimes , \oplus , \exists , \forall $. 
  \item A clause is of the form $P(x_1,...,x_n) \longleftarrow \phi (x_1,...,x_n)$, where the  atomic formula  $P(x_1,...,x_n)$ is the head, and  the formula $\phi (x_1,...,x_n)$ is the body. It is assumed that the free variables of the body are among   $x_1,...,x_n$.
  \item  A  program  is a finite set of clauses with no predicate letter appearing in the head of more than one clause (this apparent restriction causes no loss of generality \cite{fitting:bilat}).
\end{itemize}
\end{definition}

 We shall refer to such an extended logic program as a {\it Fitting program}.
Fitting also defined  the family of conventional logic programs. A {\it conventional logic program} is one whose underlying truth-value space is the bilattice $\cal FOUR$ and which does not involve $\otimes, \oplus, \forall, \cal U, \cal I$. Such programs can be written in the customary way, using commas to denote conjunction.

\section{Parameterized semantics for Fitting programs}

In the following, $\alpha \in \cal FOUR$, $\cal P$ is a Fitting program, $\cal V$($\cal B$) is the set of all valuations in $\cal B$  and Inst-$\cal P$ is the set of all ground instances of rules of $\cal P$. Some of the results in this section are inspired by \cite{fitting} which deals only with the  case  $\alpha$ = \fa.

\subsection{Immediate Consequence Operators}

First, we extend the two orderings on $\cal FOUR$ to the space of valuations~$\cal V(B)$.

\begin{definition}
Let $v_1$ and $v_2$ be in $ \cal V$($\cal B$), then
\begin{itemize}
\item $v_1 \leq_t v_2$ if and only if $v_1(A) \leq_t v_2(A)$ for all ground atoms $A$;
\item $v_1 \leq_k v_2$  if and only if $v_1(A) \leq_k v_2(A)$ for all ground atoms $A$.
\end{itemize}
\end{definition}

Under these two orderings  $ \cal V$($\cal B$) becomes a bilattice, and we have $(v \wedge w)(A) = v(A) \wedge w(A)$, and similarly  for the other operators. $ \cal V$($\cal B$) is infinitely distributive,  satisfies the infinitely interlacing conditions and has a negation and a conflation. 

The actions of valuations can be extended from atoms  to formulas as follows:
\begin{itemize}
\item$v(X \wedge Y) = v(X) \wedge v(Y)$, and similarly for the other operators,
\item $v((\exists x)\phi (x))= \bigvee_{t=closedterm} v(\phi (t)) $, and
\item$v((\forall x)\phi (x))= \bigwedge_{t=closedterm} v(\phi (t)) $.
\end{itemize}
 
The  predicate $equal(x,y)$ is a  predefined predicate defined by: for all valuations $v$,
\begin{itemize}
\item $v(equal(x,y)) = \cal T $ if $ x = y$, 
\item $v(equal(x,y)) =$\fas if $x \not = y$, and 
\item $v(\beta)=\beta$ for all $\beta$ in $\cal B$.
\end{itemize}

The following contrajoin operation  assigns  a truth value to a ground atom $A$ independently of the truth value assigned to the negation of $A$.\footnote{Our contrajoin operation is exactly the same as pseudovaluation in \cite{fitting}.
However, we prefer the term contrajoin of v and w as it is more indicative
of the fact that an operation is performed on valuations v and w.}

\begin{definition}[contrajoin]
Let v and w be in $ \cal V$($\cal B$).\\
The contrajoin of $v$ and $w$, denoted $v \bigtriangleup w$, is defined as follows: 
\begin{center}
v$\bigtriangleup$w(A)=v(A) and
v$\bigtriangleup$w($\neg$A)=$\neg$w(A), for each ground atom $A$.
\end{center}
\end{definition}

Contrajoin operations are extended to formulas by induction.
The idea is that  $v$ represents the information about $A$, and $w$ the information about $\neg A$. For example, if 
$v(innocent(John))$ = \trs$ and $ $w(innocent(John))$ = \uns then
$v \bigtriangleup w(innocent(John))$ = \tr,  whereas $\neg (v \bigtriangleup w(\neg innocent(John))$ = \un.

We can now  define a new operator $ {\Psi}_{\cal P}^\alpha $ which is inspired by \cite{fitting}. It infers new information from a contrajoin operation in a way that depends on the value of the parameter $\alpha$.

\begin{definition}
Let v and w be in $ \cal V$($\cal B$).
 The valuation $ {\Psi}_{\cal P}^\alpha (v,w)$ is defined as follows:
\begin{enumerate}
\item if the ground atom $A$ is not the head of any rule of Inst-$\cal P$, then
${\Psi}_{\cal P}^\alpha (v,w)(A)~=~\alpha $
\item if A $\leftarrow$ B occurs in  Inst-$\cal P$, then $ {\Psi}_{\cal P}^\alpha (v,w)(A) =
v \bigtriangleup w(B) $.
\end{enumerate}
\end{definition}

Clearly, the valuation $ {\Psi}_{\cal P}^\alpha (v,w)$ is in  ${\cal V(\cal B)} $, and as the interlacing conditions are satisfied by $\cal V(B)$, we can prove the following proposition.

\begin{proposition}
Let $\cal P$ be a Fitting  program.\\
(1)  Under the knowledge ordering, ${\Psi}_{\cal P}^{\alpha}$ is monotonic
 in both arguments;\\
(2)  Under the truth ordering, ${\Psi}_{\cal P}^{\alpha}$ is monotonic (and moreover continuous) in its first argument, and anti-monotonic in its second argument.
\end{proposition}
{\sc Proof.}The proof makes use of the following lemma which is an immediate consequence of the definition of contrajoin.

\begin{lemma}
Let $v_1, v_2, w_1, w_2 \in \cal V(B)$. We have:

(1) if $v_1 {\leq}_k v_2$ and $w_1 {\leq}_k w_2$, then  $v_1 \bigtriangleup w_1 {\leq}_k v_2 \bigtriangleup w_2$;

(2) if $v_1 {\leq}_t v_2$ and $w_2 {\leq}_t w_1$, then $v_1 \bigtriangleup w_1 {\leq}_t v_2 \bigtriangleup w_2$;
\end{lemma}

Now, suppose  $v_1 {\leq}_k v_2$ and let $A$ be a ground atom. We want to show  that ${\Psi}_{\cal P}^{\alpha} (v_1,w)(A) {\leq}_k {\Psi}_{\cal P}^{\alpha} (v_2,w)(A)$. If A does not occur as the head of any member of Inst-$\cal P$, then ${\Psi}_{\cal P}^{\alpha} (v_1,w)(A) = {\Psi}_{\cal P}^{\alpha} (v_2,w)(A) = \alpha$. If $A \leftarrow B \in$ Inst-$\cal P$, then ${\Psi}_{\cal P}^{\alpha} (v_1,w)(A) = v_1 \bigtriangleup w (B)$, and similarly for $v_2$, so,  by part one of the previous lemma,   ${\Psi}_{\cal P}^{\alpha} (v_1,w)(A) {\leq}_k {\Psi}_{\cal P}^{\alpha} (v_2,w)(A)$. The proof of the monotonicity in the second argument is similar. Item (2) of Proposition 1 is established by a similar argument using part 2 of the lemma.  
\cqfd

Before we continue, we recall that according to  the Knaster-Tarski theorem, a monotone operator $f$ on a complete lattice $L$ has a least fixpoint $l$ and a greatest fixpoint $g$. There are two ways of constructing these fixpoints, and each leads to a technique for proving certain properties.

Following the first way, the least fixpoint of $f$ is shown to be $\bigwedge \{x \in L | f(x) \leq x \}$. It follows that if $ f(x) \leq x$, then $l \leq x$. The greatest fixpoint of $f$ is shown to be $\bigvee \{x \in L | x \leq f(x) \}$. It follows that if $ x \leq f(x)$, then $x \leq g$.

Following the second way, one produces a (generally transfinite) sequence of members of $L$ as follows: $f_0$ is the least member of $L$. For an ordinal $n$, $f_{n+1}$ is set to be $f(f_n)$, and for a limit ordinal $\lambda$, $f_\lambda$ is set to be $\bigvee_{n < \lambda} f_n$. The limit of this sequence is the least fixpoint of $f$. This yields another method of proof: by transfinite induction. If it can be shown that each member of the  sequence $f_n$ has some property, then the least fixpoint $l$ also has the property. For the greatest fixpoint, we construct a similar sequence: $f_0$ is the greatest member of $L$. For an ordinal $n$, $f_{n+1}$ is set to be $f(f_n)$, and for a limit ordinal $\lambda$, $f_\lambda$ is set to be $\bigwedge_{n < \lambda} f_n$.

It follows from Proposition 1 that the function  $\lambda x.{{\Psi}_{\cal
P}^{\alpha}}(x,v)$ has a least fixpoint and a greatest fixpoint for each ordering. We define now  a new operator ${{\Psi '}_{\cal P}^{\alpha}}$  which  associates each  valuation $v$ with one of these fixpoints depending on the value of $\alpha$.
${{\Psi '}_{\cal P}^{\alpha}}(v)$ is the iterated fixpoint of $\lambda
x . {\Psi}_{\cal P}^{\alpha}(x,v)$ obtained from 
 an initial valuation $v_{\alpha}$ defined by: ${v_\alpha}(A)=\alpha$, for all ground atoms $ A$.

\begin{definition}

Let v be in ${\cal V(\cal B)}$. Define ${{\Psi '}_{\cal P}^{\alpha}}(v)$ to be the  limit of the sequence of valuations ($a_n$) defined as follows:
\begin{itemize}
  \item $a_0 = v_{\alpha}$;
  \item $a_n = {\Psi}_{\cal P}^{\alpha}(a_{n-1},v) $, for a successor ordinal n; 
  \item  $a_{\lambda}= \left\{
     \begin{array}{lll}
     \bigvee_{n<\lambda}{\Psi}_{\cal P}^{\alpha}(a_n,v) \mbox{ for } \alpha = \cal F\\
     \bigwedge_{n<\lambda}{\Psi}_{\cal P}^{\alpha}(a_n,v) \mbox{ for }\alpha = \cal T\\
     \bigoplus_{n<\lambda}{\Psi}_{\cal P}^{\alpha}(a_n,v) \mbox{ for }\alpha = \cal U\\
     \bigotimes_{n<\lambda}{\Psi}_{\cal P}^{\alpha}(a_n,v) \mbox{ for }\alpha = \cal I\\
     
     \end{array}
     \right.
     $ , for a limit ordinal $\lambda$.
\end{itemize}
\end{definition}

In fact, we fix the truth value of negative literals with $v$, then we compute the semantics of the positive program thus obtained (in a  similar manner  to that of  Gelfond-Lifschitz transformation).

We remark that
${{\Psi '}_{\cal
P}^{\cal U}}(v)$ is the \lfp of $\lambda x . {\Psi}_{\cal
P}^{\cal U}(x,v)$ and  ${{\Psi '}_{\cal P}^{\cal I}}$ the \gfp of $\lambda x . {\Psi}_{\cal P}^{\cal I}(x,v)$ under the knowledge ordering. ${{\Psi '}_{\cal
P}^{\cal F}}(v)$ is the \lfp of $\lambda x . {\Psi}_{\cal
P}^{\cal F}(x,v)$ and  ${{\Psi '}_{\cal P}^{\cal T}}(v)$ the \gfp of  $\lambda x . {\Psi}_{\cal P}^{\cal T}(x,v)$ under the truth ordering.

To illustrate this  definition consider the  following program $\cal P$ and let $v$ be the valuation which assigns to every ground atom the truth value \un:\\

$
\cal P
\left\{
\begin{array}{lll}
             A &  \leftarrow  &   B \wedge C\\
             D &  \leftarrow  &  \neg  B \oplus \cal T \\
             E &  \leftarrow  & A \otimes \neg D \\
             B &  \leftarrow & \cal T 
\end{array}
\right.
$
\hspace{2.5cm}
\begin{tabular}{|c|c|c|c|c|c|}
\hline
                  Atom       & A & B & C & D & E   \\
\hline
${{\Psi '}_{\cal P}^{\cal F}}(v)$ & \fa & \tr & \fa & \tr &\un \\
\hline

\end{tabular}

To compute ${{\Psi '}_{\cal P}^{\cal F}} (v)$, we first  replace all negative literals by the value \un, then we compute the least model of the positive program thus obtained (with respect to  the truth ordering)  beginning with  the valuation which assigns to every ground atom the truth value \fa.

\subsection{The family of $\alpha$-fixed models}

We recall that a valuation $v$ is a model of a program $\cal P$ \ifif for all rules $A \longleftarrow B$ in Inst-$\cal P$, $v(A) \leq_t v(B)$ \cite{fitting:wfs}. By definition of ${{\Psi}_{\cal P}^{\alpha}}$, a valuation $v$ that verifies 
${{\Psi}_{\cal P}^{\alpha}}(v,v) = v$ is a model of $\cal P$. Now, every  fixpoint $m$ of ${{\Psi '}_{\cal P}^{\alpha}}$  verifies 
${{\Psi}_{\cal P}^{\alpha}}(m,m) = {{\Psi}_{\cal P}^{\alpha}}({{\Psi '}_{\cal P}^{\alpha}}(m),m) = {{\Psi '}_{\cal P}^{\alpha}}(m) = m$, therefore $m$ is a model of $\cal P$. So we can define four new families of models that we shall call $\alpha$-fixed models.

\begin{definition}[$\alpha$-fixed models]
A valuation $v \in \cal V(B)$ is a  $\alpha$-fixed model of a program $\cal P$ \ifif $v$ is a fixpoint of ${{\Psi '}_{\cal P}^{\alpha}}$.

\end{definition}


From now on,   \fa-fixed models will be called 
$pessimistic$,  \tr-fixed models $optimistic$,  \un-fixed models  $skeptical$,  and  \inc-fixed models $inconsistent$. We can now study the family of $\alpha$-fixed models.

\begin{theorem}
 
${{\Psi '}_{\cal P}^{\alpha}}$ is monotonic under
${\leq}_k$, and anti-monotonic under $ {\leq}_t$.

\end{theorem}

{\sc Proof}. Suppose $v_1 {\leq}_k v_2$. We want to show ${{\Psi '}_{\cal P}^{\alpha}}(v_1) {\leq}_k {{\Psi '}_{\cal P}^{\alpha}}(v_2)$. Consider $\alpha = \cal F$. We define two transfinite sequences of valuations $a_n$ and $b_n$ as follows: $a_0 = b_0$ is the always $false$ valuation, the least in the truth ordering; for all $n+1$ successor ordinals, $a_{n+1} = {{\Psi}_{\cal P}^{\alpha}} (a_n,v_1)$ and $b_{n+1} = {{\Psi}_{\cal P}^{\alpha}} (b_n,v_2)$; for a limit ordinal $\lambda$, $ a_{\lambda} = {\bigvee}_{n < \lambda} a_n$ and $ b_{\lambda} = {\bigvee}_{n < \lambda} b_n$. Both sequences are increasing in the truth ordering since  ${{\Psi}_{\cal P}^{\alpha}}$ is monotonic in its first argument. The  sequence $a_n $ has  ${{\Psi '}_{\cal P}^{\alpha}} (v_1)$ as its limit, while the   sequence $b_n$ has ${{\Psi '}_{\cal P}^{\alpha}} (v_2)$ as its limit, so it is enough to establish that  $a_n {\leq}_k b_n$ for every ordinal $n$.

If  $n = 0$, $a_0 =b_0$.

Suppose  $a_n {\leq}_k b_n$. Then $a_{n+1} = {{\Psi}_{\cal P}^{\alpha}}(a_n,v_1) {\leq}_k {{\Psi}_{\cal P}^{\alpha}} (b_n,v_2) = b_{n+1}$, using the monotonicity of ${{\Psi}_{\cal P}^{\alpha}}$ in both arguments under $\leq_k$.

Finally, suppose $a_n {\leq}_k b_n$ for every  $n < \lambda$. $\cal V(B)$ satisfies the infinitary interlacing conditions so ${\bigvee}_{n < \lambda} a_n \; {\leq}_k \; {\bigvee}_{n < \lambda} b_n$.

The result for $\alpha = \cal T, U, I$ is established similarly by replacing respectively $a_0 = b_0$ (the  valuation always $false$) by the  valuation always $true$, always $unknown$, always $inconsistent$ and $\bigvee$ by $\bigwedge , \bigoplus$ and $\bigotimes$, respectively.
  
Anti-monotonicity under the truth ordering is established by a similar argument.\cqfd

Given the monotonicity of ${{\Psi '}_{\cal P}^{\alpha}}$ under the knowledge ordering and the complete lattice structure of  $\cal V(B)$ under this ordering, we can apply the Knaster-Tarski theorem, and we obtain the following result:

\begin{theorem}

${{\Psi '}_{\cal P}^{\alpha}}$ has a \lfp, denoted
$ {Fix}_{\cal U}^{\alpha} $, 
and a greatest fixpoint, denoted ${Fix}_{\cal I}^{\alpha}$,
 with respect to the knowledge ordering.\footnote{Actually, $ {Fix}_{\cal U}^{\alpha} $ and ${Fix}_{\cal I}^{\alpha}$ refer both to program $\cal P$, and should be denoted as ${Fix}_{\cal P,~U}^{\alpha}$ and ${Fix}_{\cal P,~I}^{\alpha}$, respectively. However, in order to simplify the presentation, we shall omit $\cal P$ in our notations.}

\end{theorem}

We can remark that the computation of  $Fix_{\cal U}^\alpha$, that we call $\alpha$-fixed semantics, 
is similar to the computation of the \wfss via the Gelfond-Lifschitz transformation.

Four different semantics can now be associated to a Fitting program, one for each value of $\alpha$. The following example shows how these semantics can be used  to provide different contexts, depending on the requirements.\\ 
{\bf Example.} Let $\cal P$ be the following  program:\\

$
\cal P
\left\{
\begin{array}{lll}
 Colleague(X,Y) &  \leftarrow  &  Colleague(Y,X)\\
 Colleague(a,b) &  \leftarrow  &  \cal T\\
 Colleague(a,c) &  \leftarrow  &  \cal F
\end{array}
\right.
$\\

If we have to send information to persons that we are sure to be colleagues of $b$, we have to choose the pessimistic or skeptical semantics. Indeed, under this semantics, the only person that can be proved to be a colleague of $b$ is $a$.

Now, if we want   to send information to persons that may be colleagues of $b$, then we have to choose the optimistic semantics. There are  two persons that are or may be colleagues of $b$ : $a$ and $c$. The following table summarizes the results.\\\\    
{\small
\begin{tabular}{|c|c|c|c|c|c|c|}
\hline
  Semantics & Coll(a,b) & Coll(b,a) & Coll(a,c) & Coll(c,a) & Coll(b,c) & Coll(c,b)\\
\hline
$Fix_{\cal U}^{\cal F}$  &$ \cal T$&$\cal T$ &$\cal F$ & $\cal F$ &$\cal F$ &$\cal F$\\
\hline
$Fix_{\cal U}^{\cal T}$  &$ \cal T$&$\cal T$ &$\cal F$ & $\cal F$ &$\cal T$ &$\cal T$\\
\hline
$Fix_{\cal U}^{\cal U}$  &$ \cal T$&$\cal T$ &$\cal F$ & $\cal F$ &$\cal U$ &$\cal U$\\
\hline
$Fix_{\cal U}^{\cal I}$  &$ \cal T$&$\cal T$ &$\cal F$ & $\cal F$ &$\cal I$ &$\cal I$\\
\hline
\end{tabular}
}\\\\

The behavior of ${{\Psi '}_{\cal P}^{\alpha}}$ with respect to the truth ordering is less simple because ${{\Psi '}_{\cal P}^{\alpha}}$ is anti-monotonic under this ordering. However, there is a modification of the 
Knaster-Tarski theorem dealing with precisely this case:

\begin{lemma}[\cite{yablo}]
 
Suppose that a function f is anti-monotonic on a complete lattice $\cal L$. Then there are two elements   $\mu$ and $\nu$ of $\cal L$, called extreme oscillation points of f, such that the following hold:\\ 
- $\mu$ and $\nu$ are the least and \gfp of $f^2$ (i.e. of $f$ composed with itself); \\
- $f$ oscillates between $\mu$ and  $\nu$ in the sense that $f(\mu)=\nu$ and $f(\nu)=\mu$;\\ 
- if $x$ and  $y$ are also elements  of $\cal L$ between which $f$ oscillates then  $x$ and $y$ lie between $\mu$ and $\nu$. 

\end{lemma}

 As  ${{\Psi '}_{\cal P}^{\alpha}}$ is anti-monotonic  and  $\cal V(B)$ is a complete lattice under the truth ordering, it follows that ${{\Psi '}_{\cal P}^{\alpha}}$ has two extreme oscillation points under this ordering:

\begin{proposition}

${{\Psi '}_{\cal P}^{\alpha}}$  has two extreme oscillation points denoted ${Fix}_{\cal F}^{\alpha}$ and ${Fix}_{\cal T}^{\alpha}$, with 
 ${Fix}_{\cal F}^{\alpha} \;  {\leq}_t \;  {Fix}_{\cal T}^{\alpha}$, under the truth ordering.

\end{proposition}

We can now extend the result of \cite{fitting} to any value of $\cal FOUR$.

\begin{theorem}

Let  $\cal P$ be a Fitting  program. Then we have:

\begin{center}
\begin{tabular}{lllll}

  ${Fix}_{\cal U}^{\alpha}$ &   =  &  ${Fix}_{\cal F}^{\alpha} $ & $ \otimes  $ & ${Fix}_{\cal T}^{\alpha}$ 
\\
 ${Fix}_{\cal I}^{\alpha}$ & = & ${Fix}_{\cal F}^{\alpha} $ & $ \oplus
$ & ${Fix}_{\cal T}^{\alpha}$ 
\\
 ${Fix}_{\cal F}^{\alpha}$ & = &  ${Fix}_{\cal U}^{\alpha} $ & $ \wedge
$ & ${Fix}_{\cal I}^{\alpha}$ 
\\
 ${Fix}_{\cal T}^{\alpha}$ & = & ${Fix}_{\cal U}^{\alpha} $ & $ \vee
$ & ${Fix}_{\cal I}^{\alpha}$ 
\\
\end{tabular}
\end{center}
\end{theorem}

{\sc Proof}.The proof of this theorem is given in the Appendix.\cqfd

The family of $\alpha$-fixed models of a program is bounded for each $\alpha \in \cal FOUR$ as follows:  in the knowledge ordering, all  $\alpha$-fixed models are between ${{Fix}}_{\cal U}^{\alpha}$ and ${Fix}_{\cal I}^{\alpha}$ which are the least and greatest $\alpha$-fixed models, respectively; in the truth ordering, all  $\alpha$-fixed models are between  ${Fix}_{\cal F}^{\alpha}$ and ${Fix}_{\cal T}^{\alpha}$ which are not necessarily  $\alpha$-fixed models of $\cal P$.

It is interesting to note that for $\alpha = \cal F$ the first equality of Theorem 3 relates two different definitions of the \wfs: the left-hand side, ${{Fix}}_{\cal U}^{\alpha}$, represents the definition of Przymusinski \cite{przy3} via three-valued stable models, whereas the right-hand side, ${Fix}_{\cal F}^{\alpha}  \otimes  {Fix}_{\cal T}^{\alpha}$, represents the definition of Van Gelder via alternating fixpoints \cite{gelder3}. Working with bilattices, Fitting generalized the approach of Van Gelder in \cite{fitting:wfs} and that of Przymusinski in \cite{fitting}.

\subsection{An algorithm for computing $\alpha$-fixed semantics}

In this section, all literals are ground literals.

An interpretation ${\cal M} = (T,F)$ is a pair of sets of atoms where $T$ is the set of atoms considered as true and $F$ the set  of atoms considered as false. The logical value of an atom $A$ with respect to $\cal M$ is :
\begin{itemize}
\item $\cal T$ if $A \in T$ and $A \not\in F$,
\item \fas if $A \not\in T$ and $A \in F$,
\item \uns  if $A \not\in T$ and $A \not\in F$, and 
\item \incs  if $A \in T$ and $A \in F$.
\end{itemize}

A pseudo-interpretation ${\cal J} = (T,F,T',F')$  is \
composed of four sets of atoms and assigns to every literal $L$ a logical value as follows:\\
-if $L$ is a ground atomic formula of the form $R(v_1,...,v_n)$ then its logical value with respect to $\cal J$ is the logical value of $R(v_1,...,v_n)$ with respect to the interpretation $(T,F)$;\\
-if $L$ is a ground atomic formula of the form $\neg R(v_1,...,v_n)$ then its logical value with respect to $\cal J$ is the negation of the logical value of $R(v_1,...,v_n)$ with respect to the interpretation $(T',F')$;

The logical value of a formula with respect to a pseudo-interpretation $\cal J$ is given by the logical value of its literals with respect to $\cal J$ and the truth tables  of the different operators.

The following algorithm uses a bottom-up approach to compute the $\alpha$-fixed semantics of a ground  Fitting program $\cal P$ with no function symbol over the bilattice $\cal FOUR$.\\\\
{\bf Algorithm: $\alpha$-fixed semantics}
\begin{tabbing}
111 \= 22\= 33\= 44\= 55\= 66 \=7 \kill
1. \> {\bf begin}\\
2. \> \> Res\_True := $\emptyset$;\\
3. \>\>  Res\_False := $\emptyset$;\\
4. \>\>  Tmp\_Res := (\{$ < >$ \},\{ $< >$ \});\\
5. \>\>  {\bf match} $\alpha$ {\bf with}\\
6. \>\>\> $\alpha$ = \trs -$>$ \\
   \>\>\>\> Init\_True := $\cal B_P$;\\
   \>\>\>\> Init\_False := $\emptyset$;\\
   \>\>\>\> Not\_Head\_True := \{ all atoms in $\cal B_P$ which are not  heads of any rule in $\cal P$ \};\\
   \>\>\>\> Not\_Head\_False := $\emptyset$;\\     
7. \>\>\> $\alpha$ = \fas -$>$ \\
   \>\>\>\> Init\_True := $\emptyset$;\\
   \>\>\>\> Init\_False := $\cal B_P$;\\
   \>\>\>\> Not\_Head\_True := $\emptyset$ ;\\
   \>\>\>\> Not\_Head\_False := \{ all atoms in $\cal B_P$ which are not heads of any rule in $\cal P$ \};\\     
8. \>\>\> $\alpha$ = \incs -$>$ \\
   \>\>\>\> Init\_True := $\cal B_P$;\\
   \>\>\>\> Init\_False := $\cal B_P$;\\
   \>\>\>\> Not\_Head\_True := \{ all atoms in $\cal B_P$ which are not heads of any rule in $\cal P$ \} ;\\
   \>\>\>\> Not\_Head\_False := \{ all atoms in $\cal B_P$ which are not heads of any rule in $\cal P$ \};\\     
9. \>\>\> $\alpha$ = \uns -$>$ \\
   \>\>\>\> Init\_True := $\emptyset$;\\
   \>\>\>\> Init\_False := $\emptyset$;\\
   \>\>\>\> Not\_Head\_True := $\emptyset$ ;\\
   \>\>\>\> Not\_Head\_False := $\emptyset$;\\

10. \>\>  {\bf while} Tmp\_Res $\not =$ (Res\_True,Res\_False) {\bf do}\\
11. \>\>\>  Tmp\_Res $ =$ (Res\_True,Res\_False);\\
12. \>\>\>  Iter\_True := Init\_True;\\
13. \>\>\>  Iter\_False := Init\_False;\\
14. \>\>\>  Tmp\_Iter := (\{$ < >$ \},\{$ < >$ \});\\
15.\>\>\>  {\bf while} Tmp\_Iter $\not =$ (Iter\_True,Iter\_False) {\bf do}\\
16.\>\>\>\>  Tmp\_Iter $=$ (Iter\_True,Iter\_False);\\
17.\>\>\>\>  Im\_\trs := $\emptyset$;\\
18.\>\>\>\>  Im\_\fas := $\emptyset$;\\
19.\>\>\>\>  Im\_\incs := $\emptyset$;\\
20.\>\>\>\>  {\bf for all} clauses $C$ in $\cal P$ {\bf match} the logical value $l$ of the body of $C$ \\
   \>\>\>\>  with respect to the pseudo-interpretation\\ 
   \>\>\>\>  (Iter\_True, Iter\_False, Res\_True, Res\_False)\\
   \>\>\>\> {\bf with}\\
21.\>\>\>\>\> $l = \cal T$ -$>$ Im\_\trs := Im\_\trs $\cup$ \{ head($C$)\}\\
22.\>\>\>\>\> $l = \cal F$ -$>$ Im\_\fas := Im\_\fas $\cup$ \{ head($C$)\}\\
23.\>\>\>\>\> $l = \cal I$ -$>$ Im\_\incs := Im\_\incs $\cup$ \{ head($C$)\}\\
24.\>\>\>\> {\bf end for}\\
25.\>\>\>\> Iter\_True := Im\_\trs $\cup$ Im\_\incs  $\cup $ Not\_Head\_True;\\
26.\>\>\>\> Iter\_False := Im\_\fas $\cup$ Im\_\incs $\cup $ Not\_Head\_False;\\
27.\>\>\> {\bf end while}\\
28.\>\>\> Res\_True := Res\_True $\cup$ Iter\_True;\\
29.\>\>\> Res\_False := Res\_False $\cup$ Iter\_False;\\
30.\>\> {\bf end while}\\
31.\>\> {\bf return} (Res\_True,Res\_False);\\
32.\> {\bf end.}    
\end{tabbing}

Intuitively, the assignment of the logical value $\alpha$ to  the atoms which are not  heads of any rule is done through the sets of atoms Not\_Head\_True and Not\_Head\_False. The  value of $\alpha$ also determines the initial value, (Init\_True, Init\_False) of the iterated computation of ${\Psi'}_{\cal P}^\alpha (v)$ performed  by the while loop (lines 15 to 27). Here   $v$ corresponds to the interpretation (Res\_True,Res\_False), and (Iter\_True, Iter\_False) to the value of a step of this computation. The first while loop (lines 10 to 30) calculates the  sequence of iterated  values of ${\Psi'}_{\cal P}^\alpha $  with \uns as initial value, and having $Fix^\alpha$ as  limit.

This algorithm could be easily modified in order to verify if an interpretation is a $\alpha$-fixed model of a Fitting program $\cal P$.

\section{Comparing the usual semantics of logic programs}

In this section, we compare the $\alpha$-fixed models of  conventional logic programs with the usual semantics, then we compare the different usual semantics among them.

The following theorem  states that the family of stable models is included in the family of pessimistic fixed models (thus extending  stable models from conventionnal logic programs to Fitting programs), and that the \wfss and the \kks are captured (and similarly extended) by our appproach.  

\begin{theorem}

Let $\cal P$ be a conventional logic program.\\
(1) If v is a three-valued stable model of $\cal P$, then v is a pessimistic fixed model.\\
(2)  If v is the \wfss of $\cal P$, then v =  ${Fix}_{\cal U}^{\cal F}$;\\
(3) If v is the \kks of $\cal P$, then v= ${Fix}_{\cal U}^{\cal U}$.
\end{theorem}

{\sc Proof}. The Gelfond-Lifschitz transformation $GL_{\cal P}$ is divided in two steps: firstly, it transforms the program $\cal P$ in a positive  program ${\cal P}_{/v}$ by replacing negative literals by their value in the valuation $v$; then, it applies to this program the immediate consequence operator ${{\Phi}_{\cal P}}_{/v}$. The valuation $v$ is a stable model if and only if $GL_{\cal P} (v) = v$.\\
We have  

${{\Phi}_{\cal P}}_{/v} (w) = {\Psi}_{\cal P}^{\cal F} (w,v)$,\\
so,

$lfp_t \; \lambda  w .\; {{\Phi}_{\cal P}}_{/v} (w) = lfp_t \; \lambda  w.\; {\Psi}_{\cal P}^{\cal F} (w,v)$.\\
Thus

$GL_p = {\Psi '}_{\cal P}^{\cal F}$ \\
so, if $v$ is a stable model of $\cal P$, then it is a fixpoint of  ${\Psi '}_{\cal P}^{\cal F}$ and consequently, a pessimistic fixed model of $\cal P$.\\
Thus, (1) is established and (2) is immediate with this proof because the well-founded semantics of $\cal P$ and $Fix_{\cal U}^{\cal F}$ are the least fixpoints under the truth ordering  of $GL_{\cal P}$ and ${\Psi '}_{\cal P}^{\cal F}$, respectivly.\\

Concerning (3), we have  ${\Psi}_{\cal P}^{\cal U} (v,v) = {\Phi}_{\cal P} (v)$ where ${\Phi}_{\cal P}$ is the Kripke-Kleene operator. Let $K_{\cal P}$ be the  Kripke-Kleene semantics, then we have

 $K_{\cal P} = lfp_k \; \lambda  x.\; {{\Phi}_{\cal P}} (x) = lfp_k \; \lambda  x.\; {\Psi}_{\cal P}^{\cal U} (x,x)$\\
Now,  ${Fix}^{\cal U}_{\cal U}$  is a fixpoint of $\lambda  x.\; {\Psi}_{\cal P}^{\cal U} (x,x)$,\\
so $K_{\cal P} \; {\leq}_k \; {Fix}^{\cal U}_{\cal U}$.\\
 In the other direction, we have ${\Psi '}_{\cal P}^{\cal U} (K_{\cal P}) = lfp_k(\lambda x. \; {\Psi}_{\cal P}^{\cal U}(x, K_{\cal P})$\\
Now, $K_{\cal P}$ is a fixpoint of  $\lambda x. \; {\Psi}_{\cal P}^{\cal U}(x, K_{\cal P})$, so ${\Psi '}_{\cal P}^{\cal U} (K_{\cal P}) \; {\leq}_k \; K_{\cal P}$.\\
As ${Fix}^{\cal U}_{\cal U}$ is the least fixpoint of  ${\Psi '}_{\cal P}^{\cal U}$, we have ${Fix}^{\cal U}_{\cal U} \; {\leq}_K \; K_{\cal P}$.\cqfd

It is important to recall here that, in our approach, positive and negative information are treated separately during the computation of ${Fix}^{\cal U}_{\cal U}$. This is not the case with the  computation of  \kk. Nevertheless, when we restrict our attention to conventional programs, the two methods compute the same semantics. 
Our approach unifies the computation of usual semantics, and thus allows us to compare them.

\newpage
\begin{theorem}

Let $\cal P$ be a Fitting  program. Then we have:
\begin{center}
${Fix}_{\cal U}^{\cal U}$ $ {\leq}_k$ $ {Fix}_{\cal U}^{\cal F}$  and
${Fix}_{\cal U}^{\cal U}$ $  {\leq}_k$ $  {Fix}_{\cal U}^{\cal T}$.
\end{center}  

\end{theorem}

{\sc Proof}. Let $A$ be a ground  atom. If  $A$ does not occurs as the head of any rule in Inst-$\cal P$, then

 ${\Psi}_{\cal P}^{\cal U} ({Fix}^{\cal F}_{\cal U},{Fix}^{\cal F}_{\cal U})(A) = {\cal U} \; {\leq}_k \; {Fix}^{\cal F}_{\cal U}(A)  = \cal F$.\\
 If  $A$ occurs as the head of a rule in Inst-$\cal P$, then

${\Psi}_{\cal P}^{\cal U} ({Fix}^{\cal F}_{\cal U},{Fix}^{\cal F}_{\cal U})(A) = {Fix}^{\cal F}_{\cal U}(A)$\\ because ${Fix}^{\cal F}_{\cal U}$ is a fixed model. Thus, we have 

 ${\Psi}_{\cal P}^{\cal U} ({Fix}^{\cal F}_{\cal U},{Fix}^{\cal F}_{\cal U}) \; {\leq}_k \; {Fix}^{\cal F}_{\cal U}$.\\
As ${\Psi '}_{\cal P}^{\cal U} ({Fix}^{\cal F}_{\cal U})$ is the least fixpoint of  $\lambda x. \; {\Psi}_{\cal P}^{\cal U} (x,{Fix}^{\cal F}_{\cal U})$, we have 

 ${\Psi '}_{\cal P}^{\cal U} ({Fix}^{\cal F}_{\cal U}) \; {\leq}_k \; {Fix}^{\cal F}_{\cal U}$.\\
Now, ${Fix}^{\cal U}_{\cal U}$ is the least fixpoint of ${\Psi '}_{\cal P}^{\cal U}$, so ${Fix}^{\cal U}_{\cal U} \; {\leq}_k \; {Fix}^{\cal F}_{\cal U}$.\\
Similarly, ${Fix}^{\cal U}_{\cal U} \; {\leq}_k \; {Fix}^{\cal T}_{\cal U}$.\cqfd

It follows from Theorem 5 that  the skeptical semantics gives less information than the pessimistic and optimistic semantics. From this theorem, we can infer the following result:

\begin{corollary}

Let $\cal P$ be a Fitting program. Then we have:

$${Fix}_{\cal U}^{\cal U} \; {\leq}_k \; {Fix}_{\cal U}^{\cal F} \; \otimes
\; {Fix}_{\cal U}^{\cal T}$$.

\end{corollary}
\vspace{-.6cm}
{\sc Proof}. The proof is immediate using the preceding theorem and
 interla-\\cing.\cqfd

In the previous corollary, the equality is satisfied for positive programs, but if we accept negation then it is false in general.

 This corollary suggests the possibility of defining a new semantics, namely ${Fix}_{\cal U}^{\cal F}~\otimes~{Fix}_{\cal U}^{\cal T}$, that is smaller than the pessimistic and optimistic semantics but greater than the skeptical semantics. The following example shows that this semantics can be useful in certain contexts.\\

$
\cal P :
A  \longleftarrow  B \vee \neg B 
$\\

\begin{tabular}{|c|c|c|c|c|}
\hline    
    Semantics of $\cal P$     &  ${Fix}_{\cal U}^{\cal F}$  &  ${Fix}_{\cal U}^{\cal T}$   & ${Fix}_{\cal U}^{\cal U}$ &   ${Fix}_{\cal U}^{\cal F} \; \otimes \; {Fix}_{\cal U}^{\cal T}$  \\
\hline
A  &  \tr  & \tr & \un & \tr  \\
\hline
B  &  \fa  &  \tr & \un & \un \\
\hline
 
\end{tabular}\\\\
 
The program $\cal P$ seems to assert that $A$ is always true (because it is inferred from either $B$ or $ \neg B$), and this conclusion is reached  by both the optimistic and the pessimistic semantics. However, there is no reason why we should choose between  $B$ $true$ and $B$ $false$ when we cannot assert anything about the value of $B$. It seems therefore more natural in this case to take the consensus between the pessimistic and optimistic semantics, which gives the value $unknown$ to $B$.
 
Although ${Fix}_{\cal U}^{\cal F} \; \otimes \; {Fix}_{\cal U}^{\cal T}$ seems to give an interesting new  semantics, one has to check under what conditions ${Fix}_{\cal U}^{\cal F} \; \otimes \; {Fix}_{\cal U}^{\cal T}$    is actually a model. Assuming that it is a model, we can call it the {\it consensus semantics}.

 \section{Conclusion}

We have defined  parametrized semantics for the family of Fitting programs \cite{fitting}, and an algorithm for their computation. The family of Fitting programs   is very general and includes the conventional logic programs. When we restrict the class of Fitting programs to the class of conventional logic programs, the new semantics coincide with the conventional ones. This allows us to compare  conventional semantics in this new setting in which they are embedded. It  also allows us to combine conventional semantics, and thus it suggests the possibility of defining new semantics such as the consensus semantics that we proposed in this paper.   

Extending this work to  logics with signs and annotations      is a topic for future work.
 
\appendix
\subsection*{Appendix - Proof of Theorem 3}

We need  a proposition and a few lemma to prove the next result.

\begin{lemma}
Let $x \in \cal V(B)$.~$x = (x \wedge {\cal U}) \oplus (x \vee \cal U)$ and $\cal F \otimes  T = U$.
\end{lemma}
{\sc Proof}.
 
\begin{tabular}{lll}
  $(x \wedge {\cal U}) \oplus (x \vee \cal U) $ & = & $[x \oplus (x \vee {\cal U})] \wedge [ {\cal U}\oplus (x \vee {\cal U})]$ \\
                     & = & $ [(x \oplus x) \vee (x \oplus {\cal U}) ] \wedge [({\cal U}\oplus x) \vee ({\cal U}\oplus \cal U)]$ \\
                     & = & $ [x \vee x] \wedge [x \vee \cal U]$ \\
                     & = & $ x \wedge (x \vee \cal U)$ \\
                     & = & $x$. 
\end{tabular}\\
$\cal F$ is the smallest member of $\cal V(B)$ under the truth ordering so ${\cal F} {\leq}_t \cal U$ and using the interlacing conditions, we have :
${\cal F} \otimes {\cal T} {\leq}_t {\cal U}\otimes {\cal T} = \cal U$.\\
Similarly, ${\cal U}{\leq}_t {\cal T}$, so 
${\cal U}= {\cal F} \otimes {\cal U}{\leq}_t {\cal F} \otimes {\cal T}$.

The three following equations have similar proofs: ${\cal F} \oplus {\cal T} = {\cal I}$, ${\cal U} \wedge {\cal I} = \cal F$ and ${\cal U} \vee {\cal I} = \cal T$.

\begin{lemma}
Let $a, b, c \in \cal V(B)$. If $a {\leq}_t b {\leq}_t c$, then

(1) $(a \wedge {\cal U}) \otimes (c \vee {\cal U}) \; {\leq}_k \; {\cal U}$ ;

(2) $(a \vee {\cal U}) \otimes c \; {\leq}_k \; b $;

(3) $(a \wedge {\cal U}) \otimes (c \wedge {\cal U}) \; {\leq}_k \; b$ .
\end{lemma}
{\sc Proof}. Since ${\cal U} {\leq}_k {\cal F}$, by the interlacing  conditions  $a \wedge {\cal U} \; {\leq}_k \; a \wedge {\cal F} = {\cal F}$. Similarly, $c \vee {\cal U} \; {\leq}_k \; {\cal T}$. Then by the interlacing  conditions,

 $(a \wedge {\cal U}) \otimes (c \vee {\cal U}) \; {\leq}_k \; {\cal F} \otimes {\cal T}$.

 By the precedent lemma, part 1 is established.

Then, using the hypothesis and the interlacing,

 $(a \vee {\cal U}) \otimes c \; {\leq}_k \; (a \vee b) \otimes c = b \otimes c \; {\leq}_k \; b$.

Finally, $(a \wedge {\cal U}) \otimes (c \wedge {\cal U}) \; {\leq}_k \; (a \wedge {\cal U}) \otimes (c \wedge b)  = (a \wedge {\cal U}) \otimes b \; {\leq}_k \; b$.\cqfd

Now, we can prove the result we need.

\begin{lemma}
Let $a, b, c \in \cal V(B)$. If $a {\leq}_t b {\leq}_t c$, then $a \otimes c \; {\leq}_k \; b$.
\end{lemma}
{\sc Proof}. Using the precedent lemmas and interlacing,\\
\begin{tabular}{lcl}
$a \otimes c$ & = & $ [(a \wedge {\cal U}) \oplus (a \vee {\cal U})] \otimes c$\\
              & = & $ [(a \wedge {\cal U}) \otimes c]\oplus [(a \vee {\cal U}) \otimes c]$\\
              & ${\leq}_k$ & $ [(a \wedge {\cal U}) \otimes c]\oplus b$\\
              & = & $[(a \wedge {\cal U}) \otimes ((c \wedge {\cal U}) \oplus (c \vee {\cal U}))]\oplus b$\\
              & = & $[(a \wedge {\cal U}) \otimes (c \wedge {\cal U})] \oplus [(a \wedge {\cal U}) \otimes (c \vee {\cal U})]\oplus b$\\
              &${\leq}_k$& $b \oplus {\cal U} \oplus b$\\
              & = & $b$.\\
\end{tabular}\\\\

Using a similar proof, the following can also be shown:

(1) if $a {\leq}_t b {\leq}_t c$, then $b \; {\leq}_k \; a \oplus c$;

(2) if $a {\leq}_k b {\leq}_k c$, then $a \wedge c \; {\leq}_t \; b$;

(3) if $a {\leq}_k b {\leq}_k c$, then $b \; {\leq}_t \; a \vee c$.\\

\begin{proposition}
If  $f$ is a monotone mapping on a complete lattice, then $f$ and $f^2$ have the same least and greatest fixpoints.
\end{proposition}
{\sc Proof}. Let $a$ be the least fixpoint of $f$ and let $b$ be the least fixpoint of $f^2$.

Every fixpoint of $f$ is also a fixpoint of  $f^2$ and $b$ is the least fixpoint of $f^2$, so $b \leq a$.\\
If $x$ is a fixpoint of $f^2$, then $f^2(f(x)) = f(f^2(x)) = f(x)$ so  $f(x)$is a fixpoint of  $f^2$. $b$ is the least fixpoint of $f^2$, so $f(b)$ is a fixpoint of $f^2$ and $b \leq f(b)$. By  monotonicity, $f(b) \leq f^2(b)= b$, so $b = f(b)$. Since $a$ is the least fixpoint of $f$, $a \leq b$.\cqfd

Now, we can prove the result concerning the structure of the family of $\alpha_{\cal P}$-fixed models.\\\\
{\bf Theorem 3} {\it
Let  $\cal P$ be a Fitting  program. Then we have:}
\begin{center}
\begin{tabular}{lllll}

  ${Fix}_{\cal U}^{\alpha}$ &   =  &  ${Fix}_{\cal F}^{\alpha} $ & $ \otimes  $ & ${Fix}_{\cal T}^{\alpha}$
\\
 ${Fix}_{\cal I}^{\alpha}$ & = & ${Fix}_{\cal F}^{\alpha} $ & $ \oplus
$ & ${Fix}_{\cal T}^{\alpha}$
\\
 ${Fix}_{\cal F}^{\alpha}$ & = &  ${Fix}_{\cal U}^{\alpha} $ & $ \wedge
$ & ${Fix}_{\cal I}^{\alpha}$
\\
 ${Fix}_{\cal T}^{\alpha}$ & = & ${Fix}_{\cal U}^{\alpha} $ & $ \vee
$ & ${Fix}_{\cal I}^{\alpha}$
\\

\end{tabular}
\end{center}
{\sc Proof}. the proof is separated in several parts.\\

{\it  Part 1}.
 We want to show that ${Fix}_{\cal F}^{\alpha} \otimes {Fix}_{\cal T}^{\alpha}$ and ${Fix}_{\cal F}^{\alpha} \oplus {Fix}_{\cal T}^{\alpha}$ are fixpoints of  ${\Psi '}_{\cal P}^{\alpha}$ in order to have ${Fix}_{\cal U}^{\alpha} \; {\leq}_k \; {Fix}_{\cal F}^{\alpha} \otimes {Fix}_{\cal T}^{\alpha}$ and ${Fix}_{\cal F}^{\alpha} \oplus {Fix}_{\cal T}^{\alpha}  \; {\leq}_k \; {Fix}_{{\cal I}}^{\alpha} $.\\
By monotonicity of ${\Psi '}_{\cal P}^{\alpha}$ under knowledge ordering, we have 

${\Psi '}_{\cal P}^{\alpha} ({Fix}_{\cal F}^{\alpha} \otimes {Fix}_{\cal T}^{\alpha}) \; {\leq}_k \; {\Psi '}_{\cal P}^{\alpha} ({Fix}_{\cal F}^{\alpha} ) = {Fix}_{\cal T}^{\alpha}$,
 
${\Psi '}_{\cal P}^{\alpha} ({Fix}_{\cal F}^{\alpha} \otimes {Fix}_{\cal T}^{\alpha}) \; {\leq}_k \; {\Psi '}_{\cal P}^{\alpha} ({Fix}_{\cal T}^{\alpha} ) = {Fix}_{\cal F}^{\alpha}$.

So
${\Psi '}_{\cal P}^{\alpha} ({Fix}_{\cal F}^{\alpha} \otimes \cal P$-${Fix}_{\cal T}^{\alpha}) \; {\leq}_k \; {Fix}_{\cal F}^{\alpha} \otimes {Fix}_{\cal T}^{\alpha}$.

Also, ${Fix}_{\cal F}^{\alpha} \; {\leq}_t \;  {Fix}_{\cal T}^{\alpha}$, so, by interlacing,

${Fix}_{\cal F}^{\alpha} = {Fix}_{\cal F}^{\alpha} \otimes {Fix}_{\cal F}^{\alpha} \; {\leq}_t \; {Fix}_{\cal F}^{\alpha} \otimes {Fix}_{\cal T}^{\alpha} \; {\leq}_t \; {Fix}_{\cal T}^{\alpha}\otimes {Fix}_{\cal T}^{\alpha} = {Fix}_{\cal T}^{\alpha}$,\\
and, by anti-monotonicity of ${\Psi '}_{\cal P}^{\alpha}$ under the truth ordering,

${\Psi '}_{\cal P}^{\alpha}({Fix}_{\cal T}^{\alpha}) \; {\leq}_t \; {\Psi '}_{\cal P}^{\alpha}({Fix}_{\cal T}^{\alpha}\otimes {Fix}_{\cal T}^{\alpha} )    \; {\leq}_t \; {\Psi '}_{\cal P}^{\alpha}({Fix}_{\cal F}^{\alpha})$,

so,

 ${Fix}_{\cal F}^{\alpha}  \; {\leq}_t \; {\Psi '}_{\cal P}^{\alpha}({Fix}_{\cal T}^{\alpha}\otimes {Fix}_{\cal T}^{\alpha} )    \; {\leq}_t \; {Fix}_{\cal T}^{\alpha}$.

Using the precedent lemma,

${Fix}_{\cal F}^{\alpha}\otimes {Fix}_{\cal T}^{\alpha}  \; {\leq}_t \; {\Psi '}_{\cal P}^{\alpha}({Fix}_{\cal F}^{\alpha}\otimes {Fix}_{\cal T}^{\alpha} )$.

We have shown that ${Fix}_{\cal F}^{\alpha}\otimes {Fix}_{\cal T}^{\alpha}$ is a fixed  point of ${\Psi '}_{\cal P}^{\alpha}$.

 The proof for ${Fix}_{\cal F}^{\alpha} \oplus {Fix}_{\cal T}^{\alpha}$ is dual.

${Fix}_{\cal U}^{\alpha}$ and ${Fix}_{{\cal I}}^{\alpha}$ are the least and greatest fixpoints of  ${\Psi '}_{\cal P}^{\alpha}$, so part 1 is established. \\

{\it Part 2}. We show  now the other direction.

${Fix}_{\cal F}^{\alpha}$ and ${Fix}_{\cal T}^{\alpha}$ are the two extremal oscillation points of ${\Psi '}_{\cal P}^{\alpha}$ under the truth ordering, so  

${Fix}_{\cal F}^{\alpha} \; {\leq}_t \; {Fix}_{\cal U}^{\alpha} \; {\leq}_t \; {Fix}_{\cal T}^{\alpha}$,

Thus, using the precedent lemma,  

${Fix}_{\cal F}^{\alpha}\otimes {Fix}_{\cal T}^{\alpha} \; {\leq}_t \; {Fix}_{\cal U}^{\alpha}$.\\
The proof of the other inequality is dual. The two first equality of the theorem are established.\\

{\it Part 3}. In this part, we show the last two equality.\\

${\Psi '}_{\cal P}^{\alpha}$ is monotonic under the knowledge ordering and its least and greatest fixpoints are ${Fix}_{\cal U}^{\alpha}$ and ${Fix}_{{\cal I}}^{\alpha}$. Under the knowledge ordering, $({{\Psi '}_{\cal P}^{\alpha}})^2$ is also  monotonic   and, using the precedent proposition, has the same least and greatest fixpoints.  $({{\Psi '}_{\cal P}^{\alpha}})^2$ is also monotonic under the truth ordering and its least and greatest fixpoints  under this ordering are  ${Fix}_{\cal F}^{\alpha}$ et ${Fix}_{\cal T}^{\alpha}$.

We have 

${Fix}_{\cal U}^{\alpha} \wedge {Fix}_{{\cal I}}^{\alpha} \; {\leq}_t \; {Fix}_{\cal U}^{\alpha}$

So

$({\Psi '}_{\cal P}^{\alpha})^2 ({Fix}_{\cal U}^{\alpha} \wedge {Fix}_{{\cal I}}^{\alpha}) \; {\leq}_t \; ({\Psi '}_{\cal P}^{\alpha})^2 ({Fix}_{\cal U}^{\alpha}) = {Fix}_{\cal U}^{\alpha}$

and, similarly,  $({\Psi '}_{\cal P}^{\alpha})^2 ({Fix}_{\cal U}^{\alpha} \wedge {Fix}_{{\cal I}}^{\alpha}) \; {\leq}_t \; {Fix}_{{\cal I}}^{\alpha}$.

Consequently, using the fact that  ${Fix}_{\cal F}^{\alpha}$ is the least fixpoint of  $({\Psi '}_{\cal P}^{\alpha})^2$ under the truth ordering, 

${Fix}_{\cal F}^{\alpha} \; {\leq}_t \;  {Fix}_{\cal U}^{\alpha} \wedge  {Fix}_{{\cal I}}^{\alpha}$.

Further,  ${Fix}_{\cal F}^{\alpha}$ is a fixpoint of $({\Psi '}_{\cal P}^{\alpha})^2$, and ${Fix}_{\cal U}^{\alpha}$ and ${Fix}_{{\cal I}}^{\alpha}$ are its least and greatest fixpoints, so
 
${Fix}_{\cal U}^{\alpha} \; {\leq}_k \;  {Fix}_{\cal F}^{\alpha} \; {\leq}_k \; {Fix}_{{\cal I}}^{\alpha}$

Thus, using the lemma,

${Fix}_{\cal U}^{\alpha} \wedge {Fix}_{{\cal I}}^{\alpha} \; {\leq}_t \;  {Fix}_{\cal F}^{\alpha}$.\\

The last part is dual.\cqfd


\addcontentsline{toc}{chapter}{Bibliographie}
 
\begin{thebibliography}{99}
{\footnotesize
   \bibitem{arieliavron98} {\sc Arieli}, O. and {\sc Avron}, A.,
                        {\it The Value of Four Values},
                        Artificial Intelligence 102:97-141, 1998.                \bibitem{belnap:four} {\sc Belnap}, N. D., Jr,
                        {\it A Useful Four-Valued Logic},
                        in: J. M. Dunn and G. Epstein (eds.), Modern
                        Uses of Multiple-valued Logic, D. Reichel,
                        Dordrecht, 1977.                                       \bibitem{fitting:kk}  {\sc Fitting}, M. C., 
                        {\it A Kripke/Kleene Semantics for Logic
                        Programs},
                        J. Logic Programming, 
                        2:295-312 (1985).                                       \bibitem{fitting:bilat}  {\sc Fitting}, M. C.,
                        {\it Bilattices and the Semantics of Logic Programming},
                        J. Logic Programming,
                        11:91-116 (1991).
  \bibitem{fitting:wfs}  {\sc Fitting}, M. C.,
                        {\it Well-Founded Semantics, Generalized},
                        in: v. Saraswat and K. Ueda (eds.), Logic Programming,                         Proceeding of the 1991 International Symposium,                                MIT Press, Cambridge, MA, 71-84,  1991. 

  \bibitem{fitting}  {\sc Fitting}, M. C.,
                        {\it The Family of Stable Models},
                        J. Logic Programming,
                        17:197-225 (1993).

\bibitem{fr97}  {\sc Fuhr}, N. and {\sc R\"{o}lleke}, T.,
                        {\it HySpirit -- a Probabilistic Inference Engine for                           Hypermedia Retrieval in Large Databases},
                         (1997).


  \bibitem{gl:stable}   {\sc Gelfond}, M. and {\sc Lifschitz}, V., 
                        {\it The Stable Model Semantics for Logic
                        Programming},
                        in: R. Kowalski and K. Bowen (eds.), 
                        Proceedings of the Fifth Logic Programming
                        Symposium MIT Press, Cambridge, MA,
                        978-992, 1988.
\bibitem{ginsberg:mv val}       {\sc Ginsberg}, M. L., 
                        {\it Multi-valued Logics: a Uniform Approach to 
                             Reasoning in Artificial Intelligence}, 
                        Computational Intelligence, 4:265-316, 1988.
\bibitem{ginsberg:bil}       {\sc Ginsberg}, M. L., 
                        {\it Bilattices and modal operators},
                        J. of Logic Computation, 1:41-69, 1990.   
 \bibitem{1}           {\sc Loyer}, Y., {\sc Spyratos}, N., {\sc Stamate}, D.,
                        {\it Unification des s\'emantiques usuelles de programmes                               logiques},
                        Journ\'ees Francophones de Programmation Logique et  programmation par Contraintes, Nantes, 135-150, 1998.        \bibitem{lss99}           {\sc Loyer}, Y., {\sc Spyratos}, N., {\sc Stamate},
                           D.,
                        {\it Computing and Comparing Semantics of Programs in 
                         Four-valued Logics}, in: M. Kutylowski, 
                         L. Pacholski and T. Wierzbicki (eds.),
                         Mathematical Foundations of Computer Science 
                         ($MFCS'99$), LNCS 1672, Springer Verlag, 
                         Szklarska Poreba,
                         Poland, 1999.                     
\bibitem{Mes97}  {\sc Messing}, B.,
{\it Combining knowledge with many-valued logics},
 Data \& Knowledge Engineering, 23:297-315 (1997).

 \bibitem{MPS97}
{\sc Mobasher}, B., {\sc Pigozzi}, D. and {\sc Slutzki}, G.,
{\it Multi-valued logic programming semantics: An algebraic approach}
Theoret. Comput. Sci,
 171:(1-2), 77-109, 1997.
 

\bibitem{przy1}       {\sc Przymusinski}, T. C., 
                        {\it Extended Stable Semantics for Normal and
                        Disjunctive Programs}, 
                        in D. H. D. Warren and P. Szeredi (eds.), 
                        Proceedings of the Seventh International
                        Conference on  Logic Programming,
                        MIT Press, Cambridge, MA,
                         459-477, 1990.
  \bibitem{przy3}       {\sc Przymusinski}, T. C., 
                        {\it Well-Founded Semantics Coincides with
                        Three-Valued Stable Semantics}, Fund. Inform.,
                        13:445-463, 1990.

 
\bibitem{SS97} {\sc Spyratos}, N. and {\sc Stamate}, D.,
{\it Multivalued stable semantics for databases with uncertain information},
 Information Modelling and Knowledge Bases, VIII, 129-144, 1997.


 \bibitem{gelder3}      {\sc Van Gelder},
                        {\it The Alternating Fixpoint of Logic Programs with                           Negation},
                        in: Proceedings of the Eighth  Symposium on
                        Principles of Database Systems, ACM, Philadelphia,                             1-10, 1989.  


  \bibitem{gelder2}     {\sc Van Gelder}, A., {\sc Ross}, K. A.,
                        {\sc Schlipf}, J. S.,
                        {\it The Well-Founded Semantics for General
                        Logic Programs}, J. ACM, 38:620-650, 1991.
 
   \bibitem{yablo}       {\sc Yablo}, S.,
                        {\it Truth and Reflection},
                        J. Philos. Logic, 14:297-349, 1985. 
}
\end{thebibliography}
\end{document}